\documentclass[12pt]{iopart}

\usepackage[dvipsnames]{xcolor}
\usepackage{natbib}
\usepackage{url} 

\usepackage{setspace}
\usepackage{amssymb}
\usepackage{cancel}
\usepackage[normalem]{ulem}
\usepackage{booktabs}
\usepackage{caption}

\def\vec#1{\mathchoice
 {\mbox{\boldmath $\displaystyle#1$}}
 {\mbox{\boldmath $\textstyle#1$}}
 {\mbox{\boldmath $\scriptstyle#1$}}
 {\mbox{\boldmath $\scriptstyle#1$}}}
 
\usepackage{graphicx}
\graphicspath{{Images/}}
\usepackage{comment}

\begin{document}

\title[QMD parameter optimisation for hadron therapy]{
Systematic Parameter Optimization of Quantum Molecular Dynamics Models 
Using Multi-Ion Fragmentation Data: Implications for Hadron Therapy.
}

\author{Akihiro Haga$^{1,2*}$, Yoshi-hide Sato$^{3}$, Daiyu Fujiwara$^{1}$, Dousatsu Sakata$^{4}$, Yuki Tominaga$^{1,5}$, David Bolst$^6$,  Edward C. Simpson$^7$, Susanna Guatelli$^6$}

\address{1 Department of Biological Sciences, Tokushima University, Tokushima 770-8503, Japan}
\address{2 Research Center for Nuclear Physics (RCNP), The University of Osaka, Osaka 567-0047, Japan}
\address{3 Department of Radiotherapy, Osaka University Hospital, Osaka 565-0871, Japan}
\address{4 National Metrology Institute of Japan (NMIJ), National Institute of Advanced Industrial Science and Technology (AIST), Tsukuba 305-8560, Japan}
\address{5 Division of Health Science, Graduate School of Medicine, The University of Osaka, 565-0871, Japan}    
\address{6 Centre For Medical and Radiation Physics, University of Wollongong, Wollongong NSW 2522, Australia}
\address{7 Department of Nuclear Physics and Accelerator Applications, Research School of Physics, The Australian National University, Canberra ACT 2601, Australia}

\ead{haga@tokushima-u.ac.jp}

\begin{abstract}

\noindent
Objective:
Quantum molecular dynamics (QMD) models are widely used to simulate nuclear fragmentation in hadron therapy, but their predictive accuracy depends strongly on model parameters that are often selected empirically. This study aimed to establish an optimized QMD framework for hadron-therapy applications through systematic calibration using experimental fragmentation data.

\noindent
Approach:
Three QMD parameters were optimized for the relativistic mean-field (RMF) model with the NS2 parameter set and the Skyrme model with the SLy4 and SkM* parameter sets: the wave-packet width ($L$), maximum evolution time ($T_m$), and impact-parameter envelope factor ($b_{\mathrm{env}}$). The wave-packet width was determined from experimental 
root-mean-square 
charge radii, whereas $T_m$ and $b_{\mathrm{env}}$ were parameterized as functions of incident kinetic energy and reaction-system mass and optimized using a broad set of proton- and heavy-ion-induced fragmentation measurements covering 30–400 MeV/u. Model performance was compared with the original LiQMD model as well as the Binary Cascade (BIC) and Liège Intranuclear Cascade (INCL) models.

\noindent
Main results:
The optimized $T_m$ exhibited a strong dependence on incident energy and only a weak dependence on system mass, suggesting that the optimal transition between the dynamical QMD stage and subsequent statistical de-excitation models is primarily governed by collision energy. 
In contrast, $b_{\mathrm{env}}$ showed interaction-dependent behavior, with NS2 favoring larger peripheral-collision contributions for lighter systems at low energies, whereas the Skyrme models exhibited relatively weak dependences on energy and mass. The optimized parameterizations substantially improved agreement with experimental fragmentation data and provided improved descriptions of fragment production cross sections, angular distributions, and energy distributions. The optimized Skyrme models achieved the best overall performance and outperformed the BIC and INCL models for most datasets.

\noindent
Significance:
Accurate modeling of fragmentation processes is essential for reliable calculations of secondary-particle transport, dose deposition, and linear energy transfer
distributions in hadron therapy. By systematically optimizing nuclear initialization, collision geometry, and dynamical evolution, the proposed framework provides a physically consistent description of nuclear fragmentation across multiple observables and represents a promising foundation for future Monte Carlo simulations and treatment-planning applications.
\end{abstract}

%
%
%
%
%

\section{Introduction}
Charged-particle therapy has become an established modality in cancer treatment owing to its superior dose localization compared with conventional photon radiotherapy
(\cite{Ramaekers2011, Kawashiro18, Chen2023}). 
In particular, proton and carbon-ion therapies are now widely used clinically, while emerging treatment strategies employing heavier ions, such as oxygen and neon beams, are attracting increasing attention because of their potentially enhanced biological effectiveness and therapeutic advantages for radioresistant tumors (\cite{Ebner2021, Inaniwa2021, Jeannette2025, Masuda2025}). Accurate modeling of particle interactions in matter is therefore essential for reliable dose calculation, treatment planning, range verification, and estimation of biological dose distributions in hadron therapy.

One of the major sources of uncertainty in hadron therapy arises from nuclear fragmentation reactions occurring between incident ions and human tissues (\cite{Schardt96, Haettner06, Bolst17}). These reactions generate a large number of secondary fragments, including light charged particles and residual nuclei, which contribute to out-of-field dose, distal-tail dose distributions, and secondary radiation fields. Fragment production also strongly influences linear energy transfer (LET) and microdosimetric distributions and consequently affects the estimation of relative biological effectiveness (RBE) (\cite{Kramer10, inaniwa2014nuclear, inaniwa2020nuclear, Parisi2025}). Furthermore, recent developments in range-monitoring and quality-assurance techniques based on secondary particle detection have increased the demand for highly accurate simulations of fragmentation processes (\cite{Rahmim13, Bertolli16, Chacon2024}). As multi-ion therapy becomes clinically feasible, predictive nuclear reaction models covering a broad range of projectile species and energies are becoming increasingly important.

Monte Carlo particle transport simulations are widely employed in hadron therapy research and clinical applications. Among them, the Geant4 toolkit (\cite{Allison2006, Allison2016, Agostinelli2003}) has been extensively used for medical applications such as dose calculations, detector simulations, and nuclear reaction modeling (\cite{Guatelli2021,Guatelli2025}). Several intranuclear reaction models have been implemented in Geant4, including the Binary Cascade (BIC) (\cite{Folger2004}), Liège Intranuclear Cascade (INCL) (\cite{Boudard2013}), and Quantum Molecular Dynamics (QMD) models (\cite{Sato2022}). In particular, QMD-based approaches are considered promising for the simulation of nucleus–nucleus collisions in the therapeutic energy region because they explicitly describe the many-body dynamical evolution of nucleons and naturally reproduce fragment formation processes (\cite{Sato2024, Moore2026}).

Previously, our group developed and implemented advanced QMD models in Geant4 based on both non-relativistic (\cite{Sato2022}) and relativistic (\cite{Haga2025}) effective nuclear interactions. Skyrme-type interactions, including the SLy4 and SkM* parameter sets, as well as the relativistic mean-field (RMF) with NS2 parameter set, were introduced into the QMD framework and validated against fragmentation measurements relevant to hadron therapy. These studies demonstrated that the models provide improved agreement with experimental fragmentation data in the therapeutic energy range and for low-Z target materials relevant to human tissues. However, several important model parameters governing the reaction dynamics remain insufficiently investigated.

In QMD simulations, the reaction outcome is strongly influenced by model parameters that control the spatial and temporal evolution of the system. One such parameter is the Gaussian wave-packet width of nucleons, commonly represented by the squared width parameter $L$, which determines the spatial distribution of nucleon wave packets and directly affects nuclear density profiles and fragment formation dynamics. In previous studies, $L$ was optimized using fragmentation measurements for a specific reaction system, while employing a fixed value for all nuclei (\cite{Sato2024}). Such an approach cannot accurately reproduce nuclear 
root-mean-square (RMS)
charge 
radii
over a broad mass range and may affect the initialization of nuclei and subsequent fragmentation dynamics.
Another important parameter is the maximum impact parameter used for collision sampling, which determines the contribution of peripheral reactions (\cite{Ogawa2018}) and significantly affects fragment yields and angular distributions. 
In addition, the termination time of the QMD evolution determines when the simulation is transferred to the subsequent statistical de-excitation models and therefore affects the final fragment observables.
Despite their importance, systematic investigations of these parameters across multiple projectile species and energies relevant to hadron therapy have not yet been performed.


In this study, we systematically optimize key QMD parameters using published fragmentation data for H-, C-, N-, O-, Ne-, and Mg-ion projectiles in the energy range from 30 to 400 MeV/u, which is relevant to hadron therapy.
The optimization targets include the Gaussian wave-packet width determined from experimental nuclear 
RMS
charge radii, the impact-parameter range used in reaction sampling, and the termination time of QMD dynamical evolution. Fragmentation observables, including production, differential, and double differential cross sections, are compared with extensive experimental datasets. The performance of the optimized QMD models employing SLy4, SkM*, and NS2 effective interactions is further compared with existing Geant4 reaction models, including 
LiQMD (\cite{Sato2022}), 
BIC (\cite{Folger2004}), and 
INCL (\cite{Boudard2013}).


The aim of this work is to establish a systematically optimized QMD framework for describing nuclear fragmentation across multiple projectile species and energies relevant to hadron therapy. 
The projectile and target nuclei included in the optimization were selected not only to cover species directly relevant to current particle therapy but also to extend moderately beyond this range in order to constrain the mass dependence and assess the robustness of the parameterization. 
The optimized models are intended to improve the nuclear-reaction component of Monte Carlo particle-transport simulations. More accurate prediction of fragmentation cross sections provides an important foundation for future evaluations of dose, LET, biological dose, and secondary-particle-based range-monitoring techniques, although these end-to-end clinical and treatment-monitoring observables are not directly evaluated in the present study.

\section{Materials and Methods}

\subsection{Quantum Molecular Dynamics Model}
\begin{figure}[tb]
\hspace*{-8mm}
\includegraphics[width=17cm]{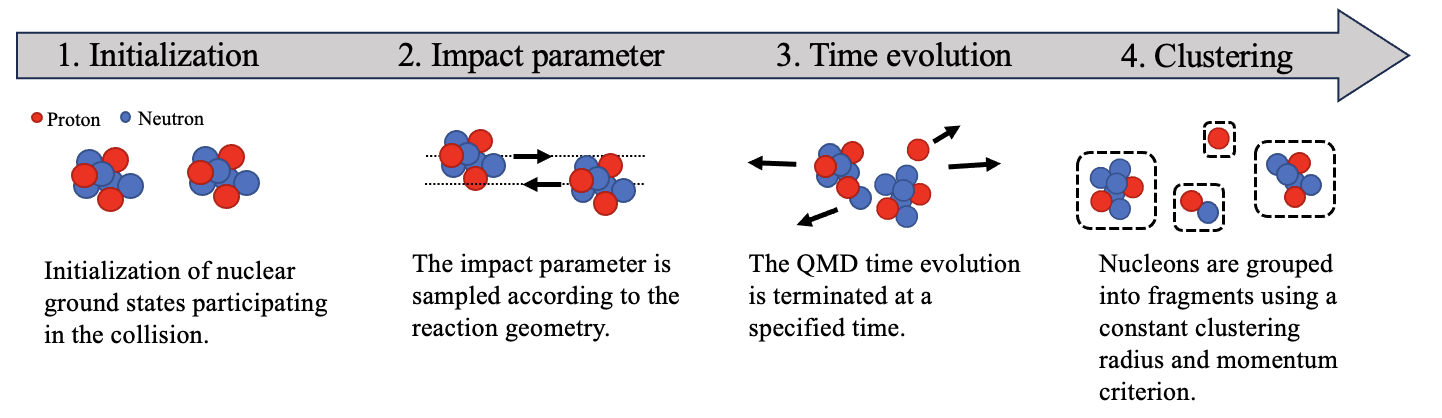}
\caption{Schematic overview of the QMD framework.
}
\label{fig1}
\end{figure}

Figure~\ref{fig1} shows the overall workflow of the QMD framework employed in this study. In the QMD model, projectile and target nuclei are initialized using Gaussian nucleon wave packets. In this study, the square of the Gaussian width, $L$, is modeled as a function of nuclear mass number in order to reproduce experimental RMS charge radii over a wide mass range. 
The collision geometry is determined by sampling the impact parameter from a uniform distribution between 0 and $b_{\mathrm{max}}$. Subsequently, the many-body dynamical evolution of nucleons is calculated using effective nuclear interactions. In this study, the Skyrme interactions with SLy4 and SkM* parameter sets (\cite{Kean20}), as well as the RMF model with NS2 parameter (\cite{Nara19, Nara2020}), are employed. The time evolution is terminated at 
a
predefined maximum evolution time, $T_m$. 
Finally, fragments are reconstructed using a clustering algorithm with a fixed spatial clustering radius $R$ and momentum conditions. 
The value of $R$, which was previously optimized using fragmentation data for a carbon-ion beam incident on a water target for the Skyrme interaction, was retained in the present study (\cite{Sato2022}). The clustering radius $R$, together with $L$, can affect fragment formation and the subsequent production of secondary particles, including prompt gamma rays and neutrons (\cite{Dedes2014}).

\subsection{Initial Nuclear State Construction and Optimization of Gaussian Wave-Packet Width}

In the QMD model, the initial nuclear state is constructed by assigning the positions and momenta of protons and neutrons before the dynamical reaction calculation. 
Geant4 currently includes two implementations of the QMD model, referred to as ``QMD'' and ``LiQMD''.
The latter was introduced in Geant4 version 11.2 and optimized for light ions relevant to heavy-ion therapy energies and nuclei (\cite{Sato2022}). The current Geant4 release (version 11.4) continues to use the same LiQMD implementation introduced in version 11.2.
In conventional QMD calculations, nucleon positions are typically sampled from phenomenological density distributions, such as 
the Woods--Saxon (\cite{Niita95}) or 
rigid-sphere distributions (\cite{Zhang20}), and 
the initial nuclear configuration was relaxed according to the QMD equations of motion so that the calculated binding energy agreed with the experimental value, following the procedure described by \cite{MaruyamaT1996}.
In LiQMD, an {\it ad hoc} alpha-cluster initialization was introduced to improve the description of light nuclei. 
The alpha-cluster-based initialization follows the procedure introduced by \cite{Sato2022} for $^{12}$C and $^{16}$O. Within each alpha cluster, four nucleons are initially arranged at the vertices of a regular tetrahedron with an inter-nucleon distance of 0.7 fm. The cluster centers are placed at predefined geometrical sites, forming an equilateral triangle for $^{12}$C and a regular tetrahedron for $^{16}$O, with an inter-cluster distance of 3.0 fm.

In the present study, this initialization was extended to nuclei with proton and neutron numbers of up to 30. 
Up to eight complete alpha clusters, corresponding to 32 nucleons, 
were constructed and placed at sites generated by extending the original tetrahedral geometry. 
Any remaining protons and neutrons were assigned to additional sites based on 
the same geometrical construction without explicitly forming further complete alpha clusters.
Random Euler rotations were applied to the internal configurations of individual alpha clusters and to the entire nuclear configuration to reduce artificial directional bias.
Although these rotations reduce the dependence on the orientation relative to the collision axis, the initial configurations retain the structural assumptions inherent in the predefined geometrical arrangements. The possible influence of these assumptions on fragmentation observables represents a limitation of the present model. For nuclei with either the proton or neutron number greater than 30, nucleon positions were sampled from a Woods--Saxon density distribution.

In the QMD framework, $k$-th nucleon wave function, $\phi_k(\vec{r})$,  is represented by a Gaussian wave packet given by
\begin{equation}
\phi_k(\vec{r}) =
\left(
\frac{1}{2\pi L}
\right)^{3/4}
\exp
\left[
-\frac{(\vec{r}-\vec{r}_k)^2}{2L}
+
\frac{i}{\hbar}(\vec{r}-\vec{r}_k)\cdot\vec{p}_k
\right],
\end{equation}
where $\vec{r}_k$ and $\vec{p}_k$ define the position and the momentum of the nucleon.
The nuclear size in this initialization scheme is controlled by the Gaussian wave-packet width, $\sqrt{L}$. 
Therefore,  $L$ was optimized so that the calculated RMS charge radii reproduce experimental values over a wide mass range. In this study, $L$ was modeled as a function of the mass number, $A$, as
\begin{equation}
L = aA^{2/3} + b,
\end{equation}
where $a$ and $b$ are model parameters.
This parameterization was based on a previous study that proposed a proportional relationship between the Gaussian wave-packet width and the nuclear radius (\cite{Wang02}).

\begin{figure}[tb]
\includegraphics[width=16cm]{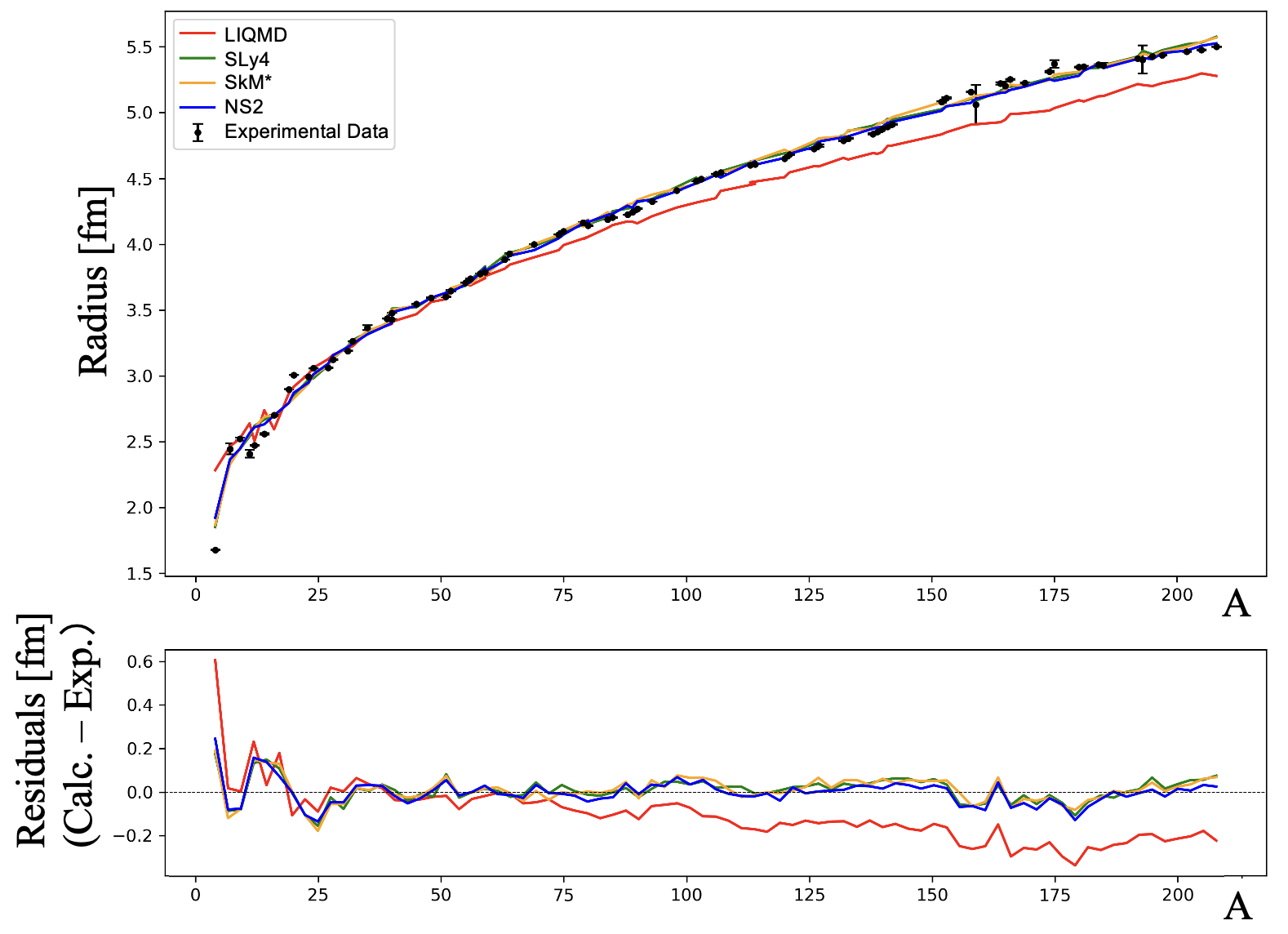}
\caption{
Comparison of calculated nuclear RMS charge radii with experimental data for stable nuclei from $^{4}$He to $^{208}$Pb. 
The upper panel shows the RMS charge radii calculated using the conventional LiQMD model with a fixed Gaussian wave-packet width and the optimized mass-dependent Gaussian widths for the SLy4, SkM*, and NS2 parameter sets. Experimental RMS charge radii obtained from the IAEA database (\cite{Angeli2013}) are also shown. 
The lower panel presents the residuals between calculated and experimental RMS charge radii. 
The conventional LiQMD model exhibits systematic deviations for both light and heavy nuclei, whereas the optimized mass-dependent Gaussian width significantly improves agreement with experimental data over the entire mass range.
}
\label{fig2}
\end{figure}

\begin{table}[tb]
\centering
\caption{Optimized parameters for the mass-dependent Gaussian wave-packet width.}
\label{tab:Lparameter}
\begin{tabular}{ccc}
\toprule
Model & $a$ & $b$ \\
\midrule
LiQMD & - & 1.26 \\
SLy4  & 0.038 & 0.89 \\
SkM*  & 0.038 & 0.89 \\
NS2   & 0.035 & 0.90 \\
\bottomrule
\end{tabular}
\end{table}

Experimental RMS charge radii for stable nuclei from $^{4}$He to $^{208}$Pb were obtained from the IAEA nuclear charge-radius database (\cite{Angeli2013}). 
The parameters $a$ and $b$ were determined independently for the Skyrme interactions SLy4 and SkM*, and for the relativistic mean-field interaction NS2, by least-squares fitting to the experimental RMS charge radii.
For comparison, the conventional LiQMD parameter set with a fixed Gaussian width ($a=0$) was also considered. The optimized parameters are summarized in Table~\ref{tab:Lparameter}.
Figure~\ref{fig2} compares the calculated RMS charge radii with experimental data for stable nuclei from $^{4}$He to $^{208}$Pb, demonstrating that the optimized mass-dependent $L$ reproduces the experimental RMS charge radii over a broad mass range.
In contrast, the conventional LiQMD with a fixed Gaussian width (\(L=1.26\) fm\(^2\)) overestimates the charge radii for lighter nuclei and underestimates them for heavier nuclei, indicating that the Gaussian width should increase with mass number to reproduce the observed nuclear size over the full mass range.
\subsection{Optimization of Impact Parameter and Maximum Evolution Time}
\label{secgeom}
In Geant4, the QMD model is invoked once after an inelastic nucleus--nucleus collision has been selected by the hadronic process. However, even when an inelastic reaction channel is selected, QMD calculations with large impact parameters may result in non-fragmenting events that effectively correspond to elastic-like scattering. In such cases, the initial QMD conditions are regenerated and the reaction calculation is repeated until a fragment-producing event is obtained. Although large impact parameters are important for describing peripheral fragmentation reactions, excessively large impact-parameter ranges significantly increase computational cost due to repeated event generation. Conversely, restricting the impact-parameter range too strongly may underestimate peripheral reactions and degrade the accuracy of fragmentation observables, particularly for light fragment production at forward angles (\cite{Ogawa2018}). Therefore, an appropriate maximum impact parameter is required to balance computational efficiency and accurate modeling of peripheral fragmentation processes.

In this study, the maximum impact parameter was defined using the reaction cross section calculated by the Geant4 hadronic interaction models. For nucleus--nucleus collisions, the reaction cross section was obtained using the Glauber--Gribov (GG) model (\cite{Grichine2009, Allison2016}), whereas the Barashenkov model (\cite{BarashenkovCompilation, Geant4PhysicsManual}) was employed for proton-induced reactions. Assuming a circular collision geometry, the maximum impact parameter was expressed as
\begin{equation}
b_{\mathrm{max}}
=
b_{\mathrm{env}}
\sqrt{\frac{\sigma_R}{\pi}},
\end{equation}
where $\sigma_R$ is the reaction cross section calculated by the corresponding Geant4 model, and $b_{\mathrm{env}}$ is an envelope factor introduced to control the contribution of peripheral collisions. Since these reaction-cross-section models already incorporate the dependence on projectile species, target species, and incident energy, the additional parameter $b_{\mathrm{env}}$ effectively represents the degree of extension beyond the nominal reaction radius required to reproduce fragmentation observables. In this study, $b_{\mathrm{env}}$ was optimized using experimental fragmentation data to account for peripheral reaction contributions while keeping the computational cost within a practically feasible range.

The maximum evolution time, $T_m$, also affects both the computational efficiency and the physical accuracy of QMD fragmentation simulations. In general, extending the QMD time evolution increases computational cost because the many-body propagation of nucleons must be calculated over a longer time interval. In addition, excessively long evolution times may induce spurious nucleon emission from excited clusters due to limitations of the QMD framework, resulting in artificial fragmentation processes after the primary reaction stage. Therefore, it is preferable to terminate the dynamical evolution during the transition from the pre-equilibrium stage toward equilibrium rather than continuing the calculation until fully stabilized final states are obtained.
On the other hand, the timescale required for fragment formation is expected to depend on the reaction system and collision violence. In particular, the equilibration process may vary according to the projectile and target mass numbers and the incident energy per nucleon. 
Accordingly, \(T_m\) was treated as a reaction-system-dependent parameter rather than as a fixed value.

The optimization of $b_{\mathrm{env}}$ and $T_m$ was performed simultaneously using published fragmentation data for H-, C-, N-, O-, Ne-, and Mg-induced reactions in the energy range from 30 to 400 MeV/u. Since both parameters are expected to depend on the reaction system and collision violence, the optimized values were parameterized as functions of the average mass number of the reaction system,
$\bar{A} = \frac{A_p + A_t}{2},$
and the incident kinetic energy per nucleon, $E_k$, where $A_p$ and $A_t$ are the projectile and target mass numbers, respectively.

The QMD simulations were performed on a discrete $b_{\mathrm{env}}$–$T_m$ grid, whereas the optimized parameters were subsequently represented by smooth sigmoid functions of the average mass number of the reaction system, $\bar{A}$, and the incident kinetic energy per nucleon, $E_k$, to provide parameter values applicable over a broad range of reaction systems relevant to hadron therapy.
QMD simulations were performed at \(b_{\mathrm{env}} = 1.0, 1.1, 1.2,\) and \(1.3\).
The envelope factor $b_{\mathrm{env}}$ was modeled within the range
$(b_{\mathrm{min}}, b_{\mathrm{max}}) = (1.0, 1.3),$
as
\begin{equation}
b_{\mathrm{env}}
=
b_{\mathrm{min}}
+
(b_{\mathrm{max}}-b_{\mathrm{min}})
\sigma(z_b),
\label{benv}
\end{equation}
where
\begin{equation}
z_b
=
w_{b,0}
+
w_{b,1}\frac{\bar{A}}{44}
+
w_{b,2}\frac{E_k}{100}
+
w_{b,3}\frac{\bar{A}}{44}\frac{E_k}{100}.
\end{equation}
Here, $\sigma(\cdot)$ is the sigmoid function and $w_{b,i}$ are fitting coefficients.
To improve numerical conditioning during optimization, \(\bar{A}\) and \(E_k\) were scaled 
down
by 44 and 100 MeV/u, respectively.
On the other hand, for $T_m$, QMD simulations were performed at
$T_m = 30, 40, 60, 80,$ and $100$~fm/$c$.
The optimized maximum evolution time was then parameterized as
\begin{equation}
T_m
=
T_{\mathrm{min}}
+
(T_{\mathrm{max}}-T_{\mathrm{min}})
\sigma(z_T),
\label{Tm}
\end{equation}
where $T_{\mathrm{min}}=30$~fm/$c$ and
\begin{equation}
z_T
=
w_{T,0}
+
w_{T,1}\frac{\bar{A}}{44}
+
w_{T,2}\frac{E_k}{100}
+
w_{T,3}\frac{\bar{A}}{44}\frac{E_k}{100}.
\end{equation}
The fitting coefficients \(w_{b,i}\) and \(w_{T,i}\), together with the upper plateau \(T_{\mathrm{max}}\), were determined simultaneously by minimizing the total loss function defined from the mean absolute errors (MAEs) between simulated and experimental fragmentation observables (presented in Section 2.4) over all reaction systems.
For reactions involving different projectile and target nuclei, the wave-packet width $L$ was set to the mass-weighted average of the optimized projectile and target values (\cite{Haga2025}).

\subsection{Experimental Fragmentation Data}
Published experimental fragmentation datasets covering $^1$H-, $^{12}$C-, $^{14}$N-, $^{16}$O-, $^{20}$Ne-, and $^{24}$Mg-induced reactions were employed for parameter optimization and evaluation of the QMD model. The selected datasets span incident energies from 30 to 400 MeV/u and cover H, C, O, Al, Ti, and Cu targets relevant to hadron therapy and radiation transport simulations.
The details of the  reactions in terms of the projectile, target, energy and cross sections are given in Appendix A.

For heavy-ion-induced reactions, experimental data reported by \cite{Dudouet2013}, \cite{Divay2017}, \cite{Zeitlin2007}, \cite{Zeitlin2011}, and \cite{Ridolfi2025} were employed. These datasets include production cross sections (PCSs), differential cross sections (DCSs, angular distributions), and double-differential cross sections (DDCSs, energy distributions at specific emission angles) of fragmentation products for C-, N-, O-, Ne-, and Mg-ion beams. In particular, fragmentation measurements at 50 and 95 MeV/u for carbon-ion beams provide detailed differential and double differential isotope fragmentation observables relevant to low-energy hadron therapy applications.

Fragment PCS for proton-induced reactions were also included in the analysis in order to extend the parameter optimization toward light-projectile reaction systems. The proton-beam datasets cover incident energies from 30 to 250 MeV and multiple target materials including C, O, Al, and Cu.

To avoid excessive bias from specific observables and experimental conditions, only selected datasets were directly included in the parameter optimization procedure. For the 50 MeV/u (\cite{Divay2017}) and 95 MeV/u (\cite{Dudouet2013}) carbon-beam experiments, DCS were used for optimization, whereas the corresponding DDCS at specific angles (3, 7, 15, 21 degrees and 4, 11, 15, and 21 degrees for 50 and 95 MeV/u carbon-ion beams, respectively) were used for consistency evaluation of the optimized parameterization. In addition, the DCS for the 400 MeV/u oxygen beam on carbon target (\cite{Ridolfi2025}) were not included in the fitting procedure and were used only for independent comparison with the optimized model.

Very heavy targets such as Sn and Pb were excluded from the optimization because the present study focused on reaction systems relevant to hadron therapy and human-body materials. 

\subsection{Benchmark Fragmentation Models}

To evaluate the effectiveness of the parameter optimization, the optimized QMD models were compared with several fragmentation models implemented in Geant4 version 11.4.0. 
The benchmark models employed in this study were the Binary Intra-Nuclear Cascade model (BIC), the Li\`ege Intra-Nuclear Cascade model (INCL), and the original LiQMD model prior to the present optimization procedure.

The BIC model describes nuclear reactions based on binary intra-nuclear cascade processes followed by pre-equilibrium and de-excitation calculations (\cite{Folger2004}). 
Because of its computational efficiency and stable performance for nucleon- and light-ion-induced reactions, BIC has been widely used in radiation transport simulations and medical physics applications.
The INCL model describes intra-nuclear cascade reactions using a dynamical cascade approach with cluster formation mechanisms and has been widely applied to fragmentation simulations in hadron-therapy studies (\cite{Boudard2013,Mancusi2014}).
However, in Geant4 version 11.4.0, the INCL model is applicable only to projectiles with mass numbers below 18 (\cite{Geant4PhysicsManual}). 
Therefore, INCL results are not presented for Ne- and Mg-ion beams in this study.
The LiQMD model corresponds to the QMD framework previously developed by our group and implemented in Geant4 since version 11.2 (\cite{Sato2022, Sato2024}). 
In the original LiQMD implementation, fixed Gaussian wave-packet widths and fixed reaction parameters were employed without systematic optimization over projectile species and incident energies. 
Therefore, comparison with the original LiQMD model allows direct evaluation of the impact of the present parameter optimization strategy.

All benchmark calculations were performed using Geant4 version 11.4.0 under the same simulation and analysis conditions for each reaction system. Because the experimental datasets provide fragmentation cross sections for specified projectile-target combinations, the simulations were performed for the corresponding projectile and target nuclei rather than by reproducing the macroscopic experimental geometry.
In Geant4, the BIC, INCL, and QMD fragmentation models are invoked after an inelastic reaction has been sampled. The fragment species, emission direction, and kinetic energy generated in each inelastic event were recorded. The resulting fragment yields and distributions were normalized using the corresponding inelastic reaction cross section calculated with the Glauber--Gribov model to obtain the PCS, DCS, and DDCS values. The same normalization procedure was applied to all fragmentation models. The same fragment-selection conditions and analysis procedures as those used for the experimental datasets were applied, including experimental energy thresholds where applicable.

\subsection{Evaluation Metric}

The agreement between simulated and experimental fragmentation observables was evaluated using the MAE. 
For each reaction condition specified by the average mass number $\bar{A}$ and incident kinetic energy per nucleon $E_k$, QMD simulations
were performed over a two-dimensional grid of $b_{\mathrm{env}}$ and $T_m$. 
Each simulation was performed with $10^4$ incident particles.
An
MAE map was then constructed on this parameter grid.

For each grid point, the MAE was calculated by averaging the absolute deviations between simulated and experimental fragmentation observables over all fragment species included in the corresponding dataset.
The correspondence between the reaction conditions and the fragmentation observables is presented in Section 3.2.
For the 50 and 95 MeV/u carbon-ion datasets, the MAE was averaged over both fragment species and measured angular bins. 
Thus, each reaction condition was represented by a two-dimensional MAE distribution as a function of $b_{\mathrm{env}}$ and $T_m$.

To compare MAE maps obtained for different reaction systems and observables on a common scale, each MAE map was normalized by its maximum value within the corresponding $b_{\mathrm{env}}$-$T_m$ grid:
\begin{equation}
\mathrm{MAE}_{\mathrm{norm}}
\left(
b_{\mathrm{env}}, T_m; \bar{A}, E_k
\right)
=
\frac{
\mathrm{MAE}
\left(
b_{\mathrm{env}}, T_m; \bar{A}, E_k
\right)
}{
\max\limits_{b_{\mathrm{env}},T_m}
\mathrm{MAE}
\left(
b_{\mathrm{env}}, T_m; \bar{A}, E_k
\right)
}.
\end{equation}
Rather than independently selecting the locally optimal grid point for each reaction condition, the normalized MAE distributions were interpolated over the sampled $b_{\mathrm{env}}$-$T_m$ range and directly used to optimize the continuous parameterization functions $b_{\mathrm{env}}(\bar A,E_k)$ and $T_m(\bar A,E_k)$. The fitting coefficients, including $w_b$ and $w_T$, were determined by minimizing the total interpolated normalized MAE over all reaction conditions.

Because the 50 and 95 MeV/u carbon-ion datasets provide angular distribution, they contain more detailed constraints on fragmentation than PCS data alone. 
However, the number of reaction conditions represented by these DCS datasets was smaller than those for the proton-induced and high-energy heavy-ion PCSs. 
Specifically, the PCS datasets comprised 56 reaction conditions in total, whereas the DCS datasets comprised 10 reaction conditions (2 energies and 5 targets). Because the MAE for each DCS condition was already averaged over the measured angular bins, each reaction condition contributed a single MAE value to the optimization. A weighting factor of 5 was therefore applied to the DCS contribution so that its effective weight (50) was comparable to that of the PCS datasets (56).
This weighting was introduced only to balance the relative influence of the different observable types.

After optimization, \(10^6\) incident particles were simulated for each reaction system corresponding to the available experimental datasets, using \(b_{\mathrm{env}}\) and \(T_m\) determined from the optimized functions of \(\bar{A}\) and \(E_k\). These high-statistics simulations were used for the final comparison with the experimental data.

\section{Results}

\subsection{Optimized parameterization of $b_{\mathrm{env}}$ and $T_m$}

\begin{figure}[tb]
\hspace*{-8mm}
\includegraphics[width=17cm]{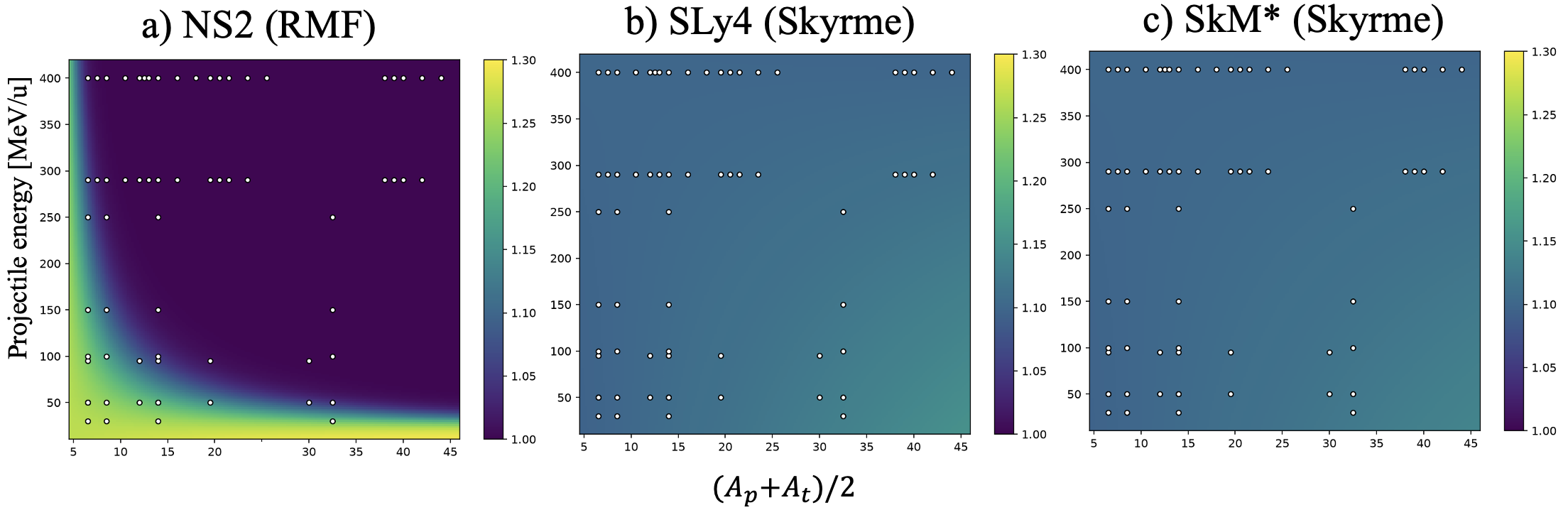}
\caption{Maps of $b_{\mathrm{env}}$ calculated using Eq.~(\ref{benv}) with the optimized parameters: 
(a) NS2 parameter set in the relativistic mean-field model, 
(b) SLy4 parameter set, and 
(c) SkM* parameter set in the non-relativistic Skyrme model.
Circles indicate the reaction systems for which experimental data are available.}
\label{fig_benv}
\end{figure}

Figure~\ref{fig_benv} shows the map of $b_{\mathrm{env}}$ calculated using Eq.~(\ref{benv}) with the optimized parameters. 
The optimized $b_{\mathrm{env}}$ maps showed different trends among the effective interactions. 
For the RMF with NS2 parameter set, larger values of approximately 1.3 were favored for lighter systems at low energies, whereas $b_{\mathrm{env}}$ approached unity as either the incident energy or the mean mass number increased. 
In contrast, the Skyrme interactions exhibited weaker variations over the investigated parameter space, with $b_{\mathrm{env}}$ mostly remaining around 1.1--1.2. 
Thus, the RMF model (NS2) showed a stronger system dependence of the optimized impact-parameter envelope, while the Skyrme model (SLy4 and SkM*) required only moderate extensions of the nominal impact-parameter range.

\begin{figure}[tb]
\hspace*{-8mm}
\includegraphics[width=17cm]{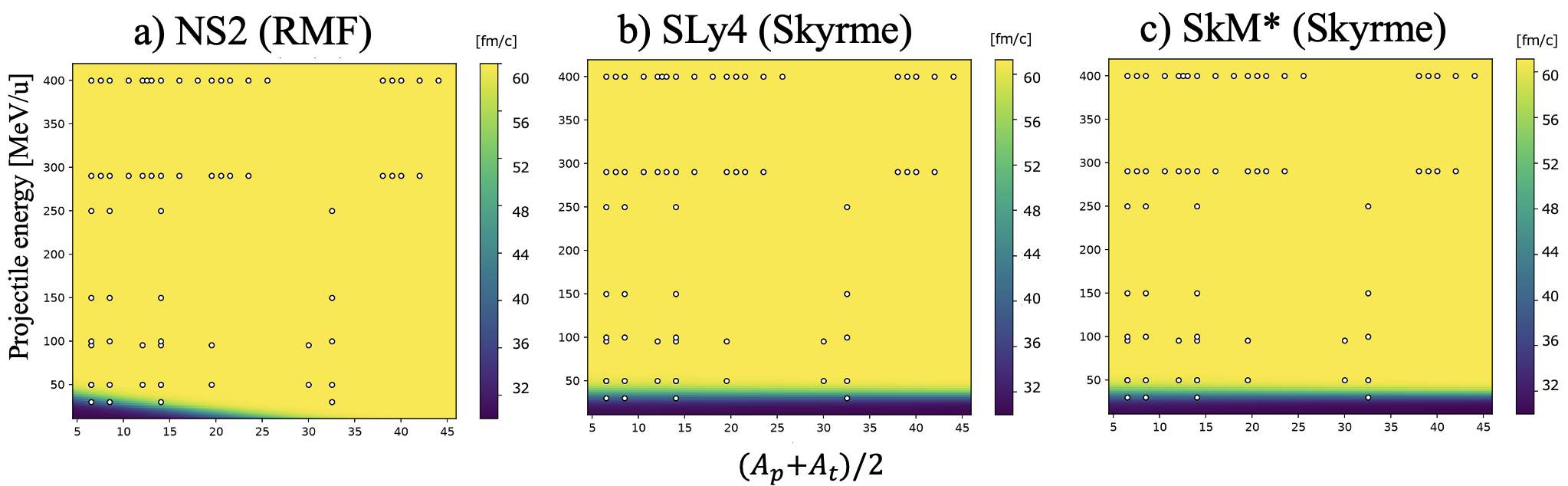}
\caption{Maps of $T_{\mathrm{m}}$ calculated using Eq.~(\ref{Tm}) with the optimized parameters for 
(a) NS2, 
(b) SLy4, and 
(c) SkM*.
Circles indicate the reaction systems for which experimental data are available.}
\label{fig_tm}
\end{figure}

Figure~\ref{fig_tm} shows the map of the optimized maximum evolution time, $T_m$, calculated using Eq.~(\ref{Tm}). 
The optimized $T_m$ showed a strong dependence on the incident kinetic energy, whereas its dependence on the mean mass number $\bar{A}$ was relatively weak within the present dataset range. 
The parameterization favored shorter evolution times of approximately 40~fm/$c$ around 30~MeV/u, 
followed by a rapid increase with incident energy.
At higher energies, \(T_m\) approached a plateau of approximately 60 fm/c for all three interactions. The NS2 interaction showed a 
stronger dependence on \(\bar{A}\) in the low-energy region than the two Skyrme interactions.

Table~\ref{tab:param} 
presents
the optimized fitting parameters for each effective interaction.
The coefficients $w_{b,i}$ and $w_{T,i}$ correspond to the sigmoid parameterizations of $b_{\mathrm{env}}$ and $T_m$, respectively, defined in Eqs.~(\ref{benv}) and (\ref{Tm}).

\begin{table}[tb]
\caption{Optimized fitting coefficients and
$T_{\mathrm{max}}$ for the proposed parameterizations of $b_{\mathrm{env}}$ and $T_m$}
\label{tab:param}
\centering
\begin{tabular}{lccc}
\hline
Parameter & NS2 & SLy4 & SkM* \\
\hline
$b_{\mathrm{env}}$\\
\hline
$w_{b,0}$ &  1.6551& -0.8659 & -0.8336 \\
$w_{b,1}$ & 4.4761 &  0.8943 &  0.6701 \\
$w_{b,2}$ &  1.8452 &  0.0432 &  0.0351 \\
$w_{b,3}$ & -20.6395 & -0.2236 & -0.1675 \\
\hline
$T_m$\\
\hline
$w_{T,0}$ & -9.1598 & -6.2712 & -6.2537 \\
$w_{T,1}$ &  10.9878 &  -2.3231 &  -3.4317 \\
$w_{T,2}$ &  21.1462 &  18.3860 &  18.3483 \\
$w_{T,3}$ &  11.2321 &   7.7437 &  11.4389 \\
$T_{\mathrm{max}}$ [fm/$c$] &  60.0371 &   61.2848 &  61.1535 \\
\hline
\end{tabular}
\end{table}

\subsection{Performance for production cross sections}

\begin{table*}[h]
\centering
\caption{Average MAE values for fragmentation observables for each dataset and model. The minimum value in each row is shown in bold.}
\label{tab:mae_summary}
\renewcommand{\arraystretch}{0.85}
\scriptsize
\small
\begin{tabular}{lrrrrrr}
\hline
Dataset & NS2 & SLy4 & SkM* & BIC & INCL & LiQMD \\
\hline
p-C & {6.11} & {\textbf{4.52}} & {4.80} & 10.20 & 9.29 & 8.48 \\
p-O &  {3.69} &  {3.95} &  {3.82} & 6.05 & \textbf{3.64} & 5.36 \\
p-Al &  {14.23} &  {10.68} &  {10.42} & 11.80 & 10.90 & \textbf{8.46} \\
p-Cu &  {8.14} &  {\textbf{4.35}} &  {4.38} & 13.20 & 14.20 & 16.20 \\
C-50-H DCS &  {\textbf{0.50}} &  {0.56} &  {0.54} & 1.07 & 1.25 & 0.57 \\
C-50-C DCS &  {\textbf{1.03}} &  {1.30} &  {1.28} & 2.00 & 2.10 & 1.97 \\
C-50-O DCS &  {\textbf{1.18}} &  {1.55} &  {1.51} & 2.23 & 2.14 & 2.35 \\
C-50-Al DCS &  {\textbf{1.52}} &  {2.13} &  {2.12} & 2.64 & 1.90 & 2.74 \\
C-50-Ti DCS &  {2.73} &  {3.59} &  {3.54} & 3.35 & \textbf{2.41} & 3.49 \\
C-95-H DCS &  {\textbf{0.28}} &  {0.30} &  {0.29} & 0.39 & 0.48 & 0.29 \\
C-95-C DCS &  {\textbf{0.82}} &  {0.85} &  {0.82} & 2.50 & 1.80 & 1.10 \\
C-95-O DCS &  {1.11} &  {1.15} &  {\textbf{1.10}} & 3.06 & 2.20 & 1.41 \\
C-95-Al DCS &  {\textbf{1.23}} &  {1.31} &  {1.24} & 3.42 & 2.03 & 1.65 \\
C-95-Ti DCS &  {\textbf{1.58}} &  {1.82} &  {1.71} & 4.17 & 2.51 & 2.20 \\
C-290-H PCS &  {\textbf{8.80}} &  {8.98} &  {8.96} & 12.77 & 13.54 & 11.12 \\
C-290-C PCS &  {18.49} &  {\textbf{14.47}} &  {15.47} & 33.06 & 29.26 & 16.00 \\
C-290-Al PCS &  {27.85} &  {\textbf{19.88}} &  {20.30} & 45.34 & 44.53 & 25.01 \\
C-290-Cu PCS &  {\textbf{29.06}} &  {31.32} &  {30.54} & 69.87 & 47.18 & 32.04 \\
C-400-H PCS &  {\textbf{12.93}} &  {13.01} &  {13.06} & 15.83 & 17.09 & 14.43 \\
C-400-C PCS &  {19.87} &  {\textbf{11.88}} &  {13.07} & 29.67 & 27.02 & 13.54 \\
C-400-Al PCS &  {29.15} &  {\textbf{24.83}} &  {25.27} & 37.51 & 48.41 & 28.27 \\
C-400-Cu PCS &  {\textbf{23.25}} &  {25.41} &  {24.86} & 45.99 & 38.73 & 25.17 \\
N-290-H PCS &  {8.79} &  {\textbf{7.33}} &  {7.61} & 13.48 & 8.64 & 7.73 \\
N-290-C PCS &  {35.37} &  {\textbf{28.82}} &  {29.78} & 50.65 & 44.56 & 31.27 \\
N-290-Al PCS &  {38.52} &  {32.54} &  {31.63} & 69.81 & 93.02 & \textbf{29.65} \\
N-290-Cu PCS &  {66.50} &  {82.00} &  {80.23} & 75.22 & 112.88 & \textbf{63.67} \\
N-400-H PCS &  {7.95} &  {\textbf{5.89}} &  {6.09} & 12.39 & 7.07 & 7.04 \\
N-400-C PCS &  {25.89} &  {15.70} &  {15.68} & 44.10 & 33.83 & \textbf{14.13} \\
N-400-Al PCS &  {46.22} &  {33.69} &  {\textbf{33.26}} & 77.29 & 87.84 & 39.44 \\
N-400-Cu PCS &  {57.29} &  {70.02} &  {68.09} & 64.53 & 94.47 & \textbf{46.03} \\
O-290-H PCS &  {6.12} &  {\textbf{5.70}} &  {5.83} & 9.49 & 14.05 & 9.78 \\
O-290-C PCS &  {13.57} &  {9.50} &  {\textbf{8.63}} & 28.29 & 29.89 & 11.41 \\
O-290-Al PCS &  {46.68} &  {33.53} &  {\textbf{33.14}} & 50.66 & 88.42 & 39.55 \\
O-290-Cu PCS &  {47.17} &  {41.79} &  {41.05} & 56.72 & 62.96 & \textbf{32.67} \\
O-400-H PCS &  {\textbf{7.40}} &  {9.12} &  {9.08} & 12.62 & 19.59 & 13.60 \\
O-400-C PCS &  {22.68} &  {11.70} &  {\textbf{11.28}} & 37.98 & 43.82 & 20.23 \\
O-400-Al PCS &  {59.65} &  {47.49} &  {\textbf{47.21}} & 83.05 & 97.80 & 58.27 \\
O-400-Cu PCS &  {39.33} &  {27.54} &  {26.20} & 40.05 & 63.81 & \textbf{26.16} \\
Ne-290-H PCS &  {9.91} &  {10.61} &  {10.57} & 9.85 & - & \textbf{8.12} \\
Ne-290-C PCS &  {\textbf{12.52}} &  {15.54} &  {15.48} & 46.91 & - & 12.60 \\
Ne-290-Al PCS &  {39.17} &  {28.13} &  {27.32} & 51.42 & - & \textbf{25.44} \\
Ne-290-Cu PCS &  {46.87} &  {54.07} &  {53.34} & 75.63 & - & \textbf{41.88} \\
Ne-400-H PCS &  {20.41} &  {18.00} &  {18.26} & 15.38 & - & \textbf{13.74} \\
Ne-400-C PCS &  {16.66} &  {\textbf{13.32}} &  {13.52} & 30.05 & - & 14.15 \\
Ne-400-Al PCS &  {44.58} &  {33.69} &  {\textbf{33.14}} & 87.02 & - & 34.02 \\
Ne-400-Cu PCS &  {54.10} &  {61.87} &  {60.67} & 90.59 & - & \textbf{48.81} \\
Mg-400-H PCS &  {32.03} &  {27.65} &  {27.35} & 22.77 & - & \textbf{20.59} \\
Mg-400-C PCS &  {\textbf{62.74}} &  {67.13} &  {66.86} & 76.35 & - & 63.33 \\
Mg-400-Al PCS &  {81.02} &  {86.82} &  {86.64} & 111.67 & - & \textbf{80.89} \\
Mg-400-Cu PCS &  {115.21} &  {116.34} &  {115.28} & 130.33 & - & \textbf{106.83} \\
\hline
\end{tabular}%
\end{table*}


Table~\ref{tab:mae_summary} summarizes the MAE values for the fragmentation observables for each dataset and model. For each heavy-ion dataset, the MAE was calculated over the included fragment species and, for DCS data, over the measured angular bins.
For the heavy-ion datasets, each reaction condition was associated with one fragmentation observable used in the optimization, whereas for the proton-induced datasets, the MAE values over the considered incident energies were summarized together.
For proton-induced reactions, the MAE values were calculated from fragment PCSs over the incident proton energy range of 30--250 MeV and are reported separately for each target material, denoted as p-C, p-O, p-Al, and p-Cu. Dataset names are given in the format projectile-energy-target-observable.
Overall, the optimized QMD models reduced the MAE for many reaction systems compared with the original LiQMD model and the cascade-based models (BIC and INCL), although the degree of improvement depended on the projectile species, target material, and incident energy.

\begin{figure}[tb]
\vspace*{-10mm}
\includegraphics[width=12.5cm]{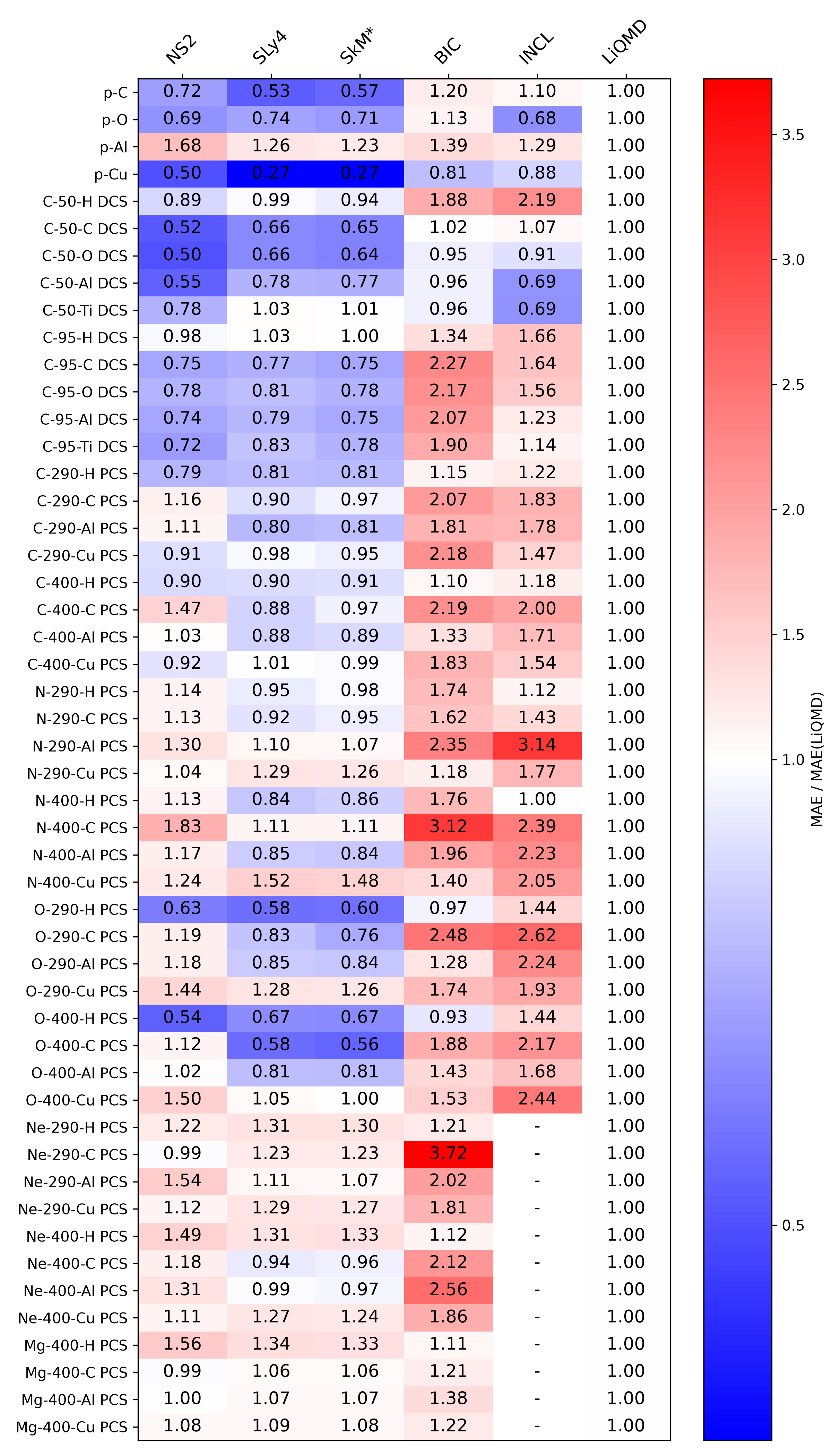}
\vspace*{-5mm}
\caption{
Heat map of the MAE ratio relative to the original LiQMD model for all fragmentation datasets. The color scale represents $\mathrm{MAE}_{\mathrm{model}}/\mathrm{MAE}_{\mathrm{LiQMD}}$, where values smaller than unity indicate improved agreement with experimental data compared with LiQMD, and values larger than unity indicate degraded agreement. Dataset labels follow the notation defined in Table~\ref{tab:mae_summary}.
}
\label{fig_mae_map}
\end{figure}

Figure~\ref{fig_mae_map} shows the ratio of the MAE for each model to that of the original LiQMD model. Values smaller than unity indicate an improvement over LiQMD, whereas values larger than unity indicate a degradation in agreement with the experimental data. Overall, the optimized QMD models showed substantial improvements for reaction systems involving light projectiles and targets. In particular, noticeable reductions in MAE were observed for proton-induced reactions and for light-ion fragmentation datasets. As the projectile and target masses increased, the performance of the optimized models became comparable to that of LiQMD, and in some cases slightly larger MAE values were obtained. However, these differences were generally modest compared with the improvements observed for lighter systems.

Compared with the cascade-based models, all QMD-based models generally provided substantially better agreement with the experimental fragmentation data.
Furthermore, the present parameter optimization further reduced the MAEs for many reaction systems, particularly for light and intermediate-mass projectiles and targets. 
However, for several reaction systems involving relatively heavy projectiles, the conventional LiQMD model provided better agreement than the optimized models. This may indicate that, for these systems, the modification of the Gaussian wave-packet width \(L\), which was determined from ground-state nuclear-size properties rather than fragmentation observables, does not necessarily improve the description of the reaction dynamics. In addition, the NS2 interaction showed somewhat larger variations in performance than the two Skyrme interactions, suggesting a greater sensitivity of the RMF-based model to the optimized reaction parameters. This model dependence is discussed further in the Discussion section.
Overall, the optimized parameterization improves the predictive capability of the QMD models for many fragmentation reactions relevant to hadron-therapy applications.

\clearpage
\subsection{Consistency evaluation using additional fragmentation observables}

\begin{figure}[tb]
\includegraphics[width=13cm]{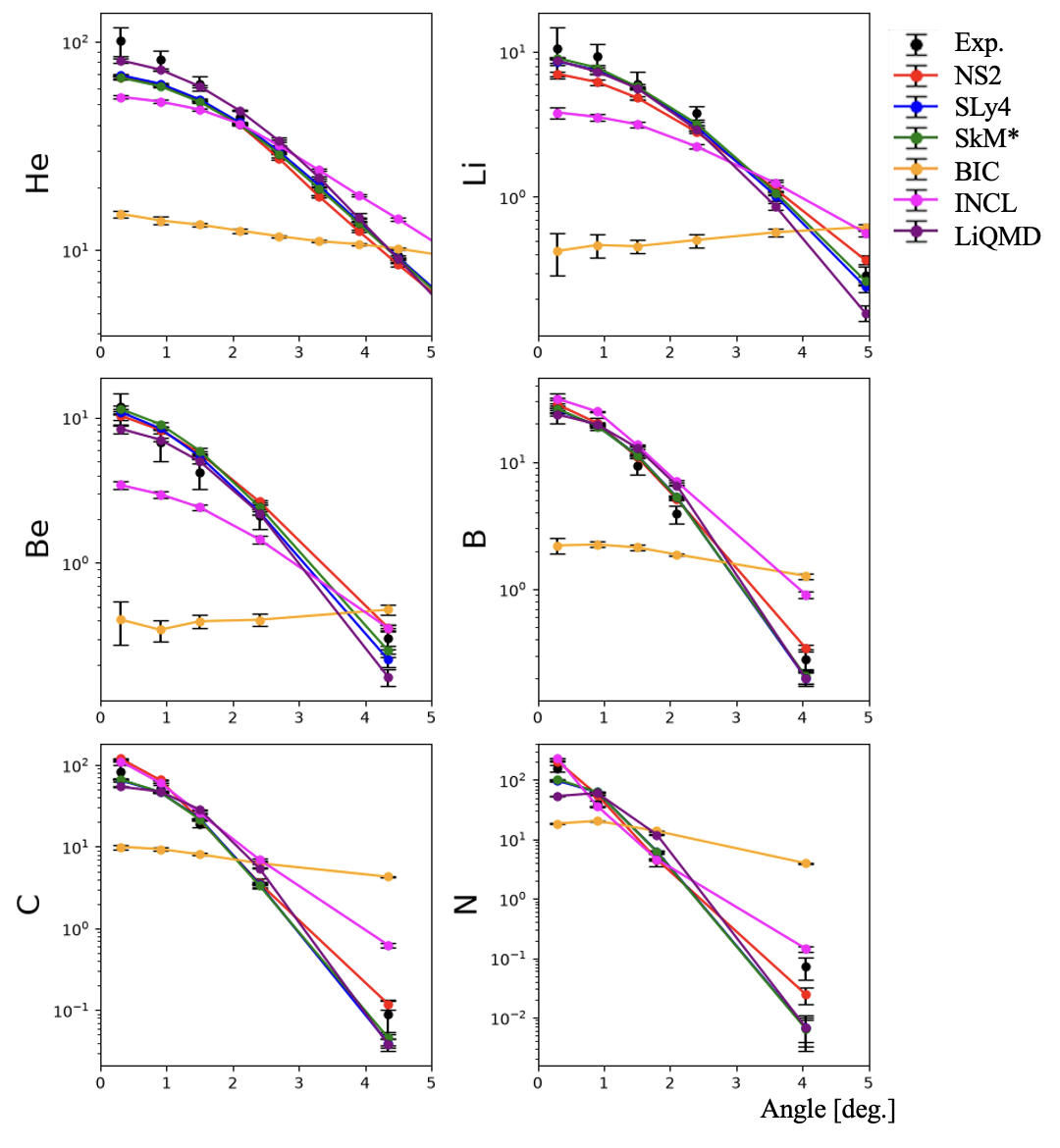}
\caption{
Angular distributions of He, Li, Be, B, C, and N fragments produced in the $^{16}$O + C reaction at 400 MeV/u. Experimental data (black circles) are compared with the predictions of the optimized QMD models (NS2, SLy4, and SkM*), the original LiQMD model, and the cascade-based models (BIC and INCL). 
}
\label{fig_O400_ang}
\end{figure}

To evaluate the consistency of the optimized parameterization beyond the datasets used for parameter optimization, additional fragmentation observables were analyzed. Figure~\ref{fig_O400_ang} compares the angular distributions of He, Li, Be, B, C, and N fragments produced in the $^{16}$O + C reaction at 400 MeV/u (\cite{Ridolfi2025}). These angular distributions were not included in the optimization procedure and therefore provide an additional consistency evaluation of the optimized models.

Table~\ref{tab:O400_angle} summarizes the average MAE and $\chi^2$ error for the measured angular distributions. Among the QMD-based models, the optimized QMD model with SLy4 and SkM* parameter sets both achieved the lowest MAE values of 5.7, followed by NS2 (6.3) and LiQMD (6.9). In contrast, the BIC model showed substantially larger discrepancies, with an average MAE of 20.1 and a $\chi^2$ error exceeding 150,
consistent with the findings reported by \cite{Ridolfi2025}. 
The INCL model provided a reasonable description of the angular distributions, while its MAE was larger than those of all QMD models. The corresponding $\chi^2$ errors showed a similar trend. 

\begin{table}[tb]
\centering
\caption{Average MAE and $\chi^2$ error for the angular distributions of He, Li, Be, B, C, and N fragments in the $^{16}$O + C reaction at 400 MeV/u. This dataset was not included in the parameter optimization.}
\label{tab:O400_angle}
\begin{tabular}{lrrrrrr}
\hline
Metric & NS2 & SLy4 & SkM* & BIC & INCL & LiQMD \\
\hline
MAE & 6.3 & \textbf{5.7} & \textbf{5.7} & 20.1 & 8.4 & 6.9
\\
$\chi^2$ error & 13.8 & 13.1 & \textbf{13.0} & 155.1 & 25.5 & 24.5 \\
\hline
\end{tabular}
\end{table}

Finally, an additional consistency test was performed using fragment energy distributions measured for carbon-ion beams at 50 and 95 MeV/u incident on H, C, O, Al, and Ti targets. For the 50 MeV/u datasets, energy distributions measured at 3°, 7°, 15°, and 21° were analyzed, whereas for the 95 MeV/u datasets, angles of 4°, 11°, 15°, and 21° were used. 
For each reaction system, measurement angle, and fragment species, the MAE of the energy distribution was calculated.
The consistency was quantified by the improvement ratio, defined as the fraction of fragment energy distributions for which the MAE was smaller than that obtained with the original LiQMD model.

\begin{figure}[tb]
\includegraphics[width=16cm]{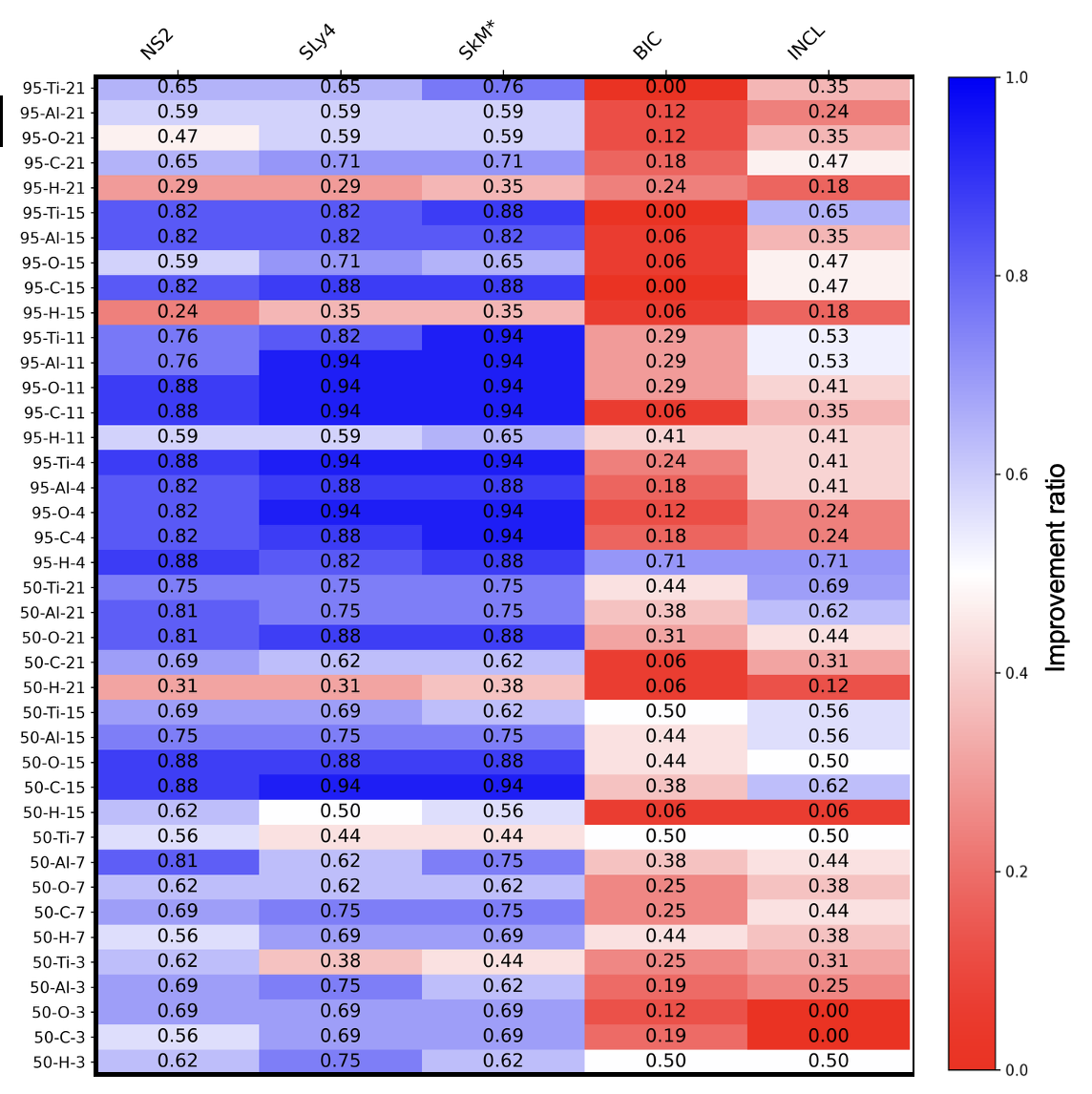}
\caption{
Improvement ratio for fragment energy distributions relative to the original LiQMD model for carbon-ion beams at 50 and 95 MeV/u incident on H, C, O, Al, and Ti targets. 
Dataset labels follow the format energy--target--angle. 
For each reaction system, measurement angle, and fragment species, the MAE of the energy distribution was compared with that obtained using LiQMD. The improvement ratio represents the fraction of fragment energy distributions for which a given model achieved a smaller MAE than LiQMD. Values approaching unity indicate that the model improved the agreement with experimental data for most fragment energy distributions, whereas values approaching zero indicate little or no improvement. The selected angles were 3°, 7°, 15°, and 21° for 50 MeV/u and 4°, 11°, 15°, and 21° for 95 MeV/u.
}
\label{fig_improvement_ratio}
\end{figure}

To compare the degree of improvement across reaction systems with different cross-section scales, we introduced an improvement ratio relative to the original LiQMD model. This dimensionless quantity represents the fraction of evaluated fragment species for which the MAE of the energy distribution was smaller than that obtained with the original LiQMD model. Thus, the improvement ratio ranges from 0 to 1, with larger values indicating that a greater fraction of the evaluated fragment energy distributions showed improved agreement relative to the original LiQMD model. This fraction-based metric facilitates comparison across different projectile-target systems and observation angles.
Figure~\ref{fig_improvement_ratio} summarizes the improvement ratio for the evaluated fragment energy distributions. 
The optimized QMD model with SLy4 and SkM* parameter sets improved the MAE for approximately 70--100\% of the evaluated fragment energy distributions in many reaction systems. The NS2 also exhibited substantial improvements over the original LiQMD model, although its improvement ratios were generally lower than those of the Skyrme models. These results demonstrate that the proposed parameter optimization improves not only fragment PCS and DCS but also fragment DDCS, providing additional evidence for the robustness of the optimized QMD framework. Notably, BIC yielded substantially lower ratios for most datasets, whereas INCL generally produced intermediate ratios. Both models were outperformed by the QMD models for the majority of the evaluated reaction systems.

\section{Discussion}
One of the most notable findings of the present study is the strong dependence of the optimized maximum evolution time, $T_m$, on the incident kinetic energy. 
The optimized parameterization favored evolution times of approximately 40 fm/$c$ for reactions around 30 MeV/u, 
and 60 fm/$c$ 
for the higher-energy datasets,
except for NS2 (RMF) at $\bar{A}>20$. 
In contrast, 
a negligible
dependence on the mean mass number $\bar{A}$ was observed.
This result suggests that the effective propagation time required to reproduce experimental fragmentation observables is governed primarily by the reaction energy rather than by the size of the reaction system. In QMD models, fragment formation emerges dynamically through nucleon--nucleon collisions and mean-field interactions followed by clusterization. As the incident energy increases, a larger number of nucleons participate in the reaction dynamics and secondary interactions become more important, potentially extending the duration of the fragmentation process. 
Consequently, terminating the calculation too early may lead to an incomplete description of the dynamical fragmentation process. 
On the other hand, excessively long QMD evolution may shift part of the de-excitation dynamics from the statistical models to the QMD stage. 
Therefore, the optimized values of $T_m$ may be interpreted not simply as a characteristic fragment formation time, but rather as the optimal transition point between the dynamical QMD stage and the subsequent statistical de-excitation models. 
The observed energy dependence suggests that higher-energy reactions require a longer dynamical evolution before the statistical treatment becomes appropriate.

In contrast to the present QMD results, the stopping time employed in the INCL model includes an explicit dependence on the system mass (\cite{Geant4PhysicsManual}). This difference likely reflects the distinct physical pictures underlying the two approaches. In INCL, the reaction is described as an intra-nuclear cascade that terminates once the cascade process has effectively ceased, leading naturally to a dependence on the size of the nucleus. The present optimization indicates that, within the QMD framework, the characteristic evolution time is instead governed primarily by the reaction energy, which controls the degree of excitation and the duration of fragment formation.

The behavior of $b_{\mathrm{env}}$ was less systematic than that of $T_m$. While the RMF model with the NS2 parameter set favored larger values for lighter systems at low energies, the Skyrme interactions exhibited only weak dependences on both system mass and incident energy. 

The individual coefficients of the sigmoid parameterization should not be interpreted as having direct physical significance; rather, the relevant quantity is the overall dependence of $b_{\mathrm{env}}$ on the reaction conditions.
One possible explanation for the NS2 trend is related to the geometrical importance of peripheral collisions
and to the different dynamical response of the RMF interaction. In the RMF framework, the nuclear mean field results from the cancellation between large attractive scalar and repulsive vector potentials, and its density-dependent evolution can differ from that of the nonrelativistic Skyrme interactions. 
This difference may increase the sensitivity of peripheral collisions to the reaction dynamics in NS2.
In addition, for a given increase in the effective reaction radius, the relative contribution of the peripheral region is larger for smaller nuclei, which is qualitatively consistent with the trend observed for NS2. On the other hand, it should be noted that the baseline reaction radius is determined from reaction-cross-section models, namely the Glauber–Gribov model for heavy ions and the Barashenkov model for proton-induced reactions, both of which already include energy and mass dependences. Therefore, the relatively small variations of the optimized $b_{\mathrm{env}}$ values for the Skyrme interactions suggest that the underlying reaction-cross-section models provide a reasonable first-order description of the collision geometry, while modest adjustments of the peripheral region are sufficient to improve the agreement with experimental fragmentation observables.

For carbon-ion beams at 50, 95, 290, and 400 MeV/u, the summed MAE over the evaluated datasets was reduced by approximately 10\% for the optimized Skyrme-based QMD models compared with the original LiQMD model. For the fragment energy-distribution datasets at 50 and 95 MeV/u, the reduction was approximately 18-20\%, and improved agreement relative to LiQMD was obtained for approximately 70-100\% of the evaluated fragment species in many cases. Since fragment production and the angular and energy distributions of secondary charged particles contribute to the spatial spread of deposited dose, these improvements may influence subsequent dose calculations. However, the present study was limited to the evaluation of fragmentation cross sections, and the quantitative impact on dose or LET distributions was not directly assessed.

The additional angular- and energy-distribution analyses provide a useful consistency check of the optimized parameterization. 
The optimized parameterizations were able to reproduce not only the integrated fragmentation observables used in the optimization but also the detailed angular and energy distributions with improved accuracy. 
This result suggests that the optimized parameters provide a more self-consistent description of fragmentation processes over multiple observables.

This consistency is relevant to hadron-therapy simulations. While fragment production cross sections determine the overall abundance of secondary particles, their angular and energy distributions govern the subsequent transport and energy deposition of the fragments. Accurate modeling of these observables therefore provides an important nuclear-reaction basis for subsequent calculations of dose, LET, and biologically weighted dose distributions. However, the improved agreement of the fragmentation observables demonstrated in the present study does not by itself establish improved accuracy of these end-to-end clinical quantities. Their predictions are also affected by the statistical de-excitation process, secondary-particle transport, and treatment geometry. Similarly, treatment-monitoring observables based on positron-emitting nuclei or prompt gamma rays require dedicated evaluation of isotope-resolved production channels and, for comparison with measured signals, particle transport and detector response. These clinical and treatment-monitoring observables were not directly evaluated in the present study and remain important subjects for future investigation.

Several limitations of the present study should be acknowledged. 
First,
different mass dependences of the optimized $b_{\mathrm{env}}$ parameter were observed for the NS2 and Skyrme interactions. Although these differences may reflect variations in the underlying effective interactions and their influence on nuclear surface properties and cluster formation dynamics, the present dataset does not allow these effects to be disentangled. Additional low-energy heavy-ion fragmentation data would be valuable for clarifying the physical origin of the optimized parameterizations. 
Second, the cluster-recognition distance $R$ was fixed at 4 fm for all reaction systems. In our previous optimization, the optimal value for the SkM* interaction was 3.99 fm, whereas somewhat larger values were favored for SLy4 depending on the objective considered; however, $R=4$ fm remained a reasonable common choice within the trade-offs of the multi-objective optimization (\cite{Sato2022}). Because $R$ affects the primary fragments transferred to the statistical de-excitation model, its value may also influence the predicted fragment and secondary-particle yields, including prompt-$\gamma$ and neutron production (\cite{Dedes2014}). The possible dependence of the optimal $R$ on the interaction model and projectile–target system was not investigated in the present study, and the use of a fixed common value should therefore be regarded as a limitation.
Finally, the relative contributions of the three optimized components ($L$, $b_{\mathrm{env}}$, and $T_m$) were not quantified in the present study. While the latter two parameters were optimized directly using fragmentation observables, preliminary investigations suggest that the optimization of $L$ may also contribute substantially to the improved agreement with experimental data through its influence on the initial nuclear density distributions and subsequent fragmentation dynamics. A dedicated sensitivity analysis would therefore be valuable for disentangling the individual roles of these parameters.

\section{Conclusion}
This study presented a systematic optimization of key parameters in QMD fragmentation models for hadron-therapy applications using a comprehensive set of experimental fragmentation data, including H-, C-, N-, O-, Ne-, and Mg-induced reactions. 
The wave-packet width $L$, maximum evolution time $T_m$, and impact-parameter envelope factor $b_{\mathrm{env}}$ were optimized for the RMF model (NS2 parameter set) and Skyrme models (SLy4, and SkM* parameter sets). While $L$ was determined from the reproduction of experimental RMS charge radii, $T_m$ and $b_{\mathrm{env}}$ were parameterized as functions of the incident kinetic energy and reaction-system mass and optimized using fragmentation observables.

The optimized $T_m$ exhibited a strong dependence on incident energy and only a weak dependence on system mass, suggesting that the transition between the dynamical QMD stage and the subsequent statistical de-excitation stage is primarily governed by the reaction energy. In contrast, the optimized $b_{\mathrm{env}}$ showed a more complex behavior, with the RMF model with NS2 parameter set favoring larger values for lighter systems at low energies, while the Skyrme interactions exhibited relatively weak dependences on both energy and mass. 
Given the observed model dependence of \(b_{\mathrm{env}}\), a dedicated validation of peripheral fragmentation channels would be valuable. In particular, reactions involving the removal of one or two nucleons from the projectile or target may be especially relevant because they can produce positron-emitting isotopes used for range verification in hadron therapy. A systematic evaluation of such peripheral reaction channels and isotope-resolved yields therefore represents an important subject for future work.

Compared with the original LiQMD model and the cascade-based BIC and INCL models, the optimized QMD parameterizations achieved substantially improved agreement with experimental fragmentation data over a wide range of reaction systems. In addition to fragment production cross sections, the optimized models provided improved descriptions of angular and energy distributions, demonstrating a consistent representation of multiple fragmentation observables.

Accurate prediction of fragmentation processes is essential for reliable calculations of secondary-particle transport, dose deposition, LET distributions, and biologically weighted dose quantities in particle therapy. The present results suggest that the optimized QMD framework provides a robust and physically consistent description of nuclear fragmentation and represents a promising approach for future Monte Carlo simulations and treatment-planning applications in hadron therapy.

\ack
This study has been supported by 
JSPS KAKENHI (Grant No. 26K10563).


\appendix
\renewcommand{\thetable}{A\arabic{table}}
\setcounter{table}{0}
\renewcommand{\thefigure}{A\arabic{figure}}
\setcounter{figure}{0}
\section{Experimental Fragmentation Datasets and Reference Cross Sections}
This appendix summarizes the experimental fragmentation datasets employed in the present study. 
Table~A1 lists the heavy-ion-induced fragmentation datasets used for parameter optimization and model evaluation.
For the proton-induced production cross sections, experimental data were compiled from various published measurements. 
For the carbon target, data were taken from Refs.~\cite{
Dickson1951_Be7_C,
Brun1962_ProtonBeamIntensity,
Gauvin1962_AlphaSpallation,
Honda1964_Spallation,
Measday1966_C11,
Davids1970_LiBeB_C12,
Fontes1971_LiBe_C12,
Fontes1977_B10B11_C12,
Vdovin1979_50MeVProtonsCNO,
Aleksandrov1990_Be7LightNuclei,
Akagi2013_PositronEmitters,
Matsushita2016_TargetFragmentationCarbon,
Horst2019_PETCrossSections}.
For the oxygen target, data were taken from Refs.~\cite{
Dittrich1990_Be10Al26,
Sisterson1997_LunarRocks,
Aleksandrov1990_Be7LightNuclei,
Michel1997_ResidualNuclides,
Bimbot1971_SpallationLightNuclei,
Valentin1963_LightNuclei155MeV,
Foley1962_GammaRadiationO16F19,
Valentin1965_MediumEnergyLightNuclei,
Akagi2013_PositronEmitters,
Masuda2018_CherenkovCrossSections,
Horst2019_PETCrossSections}.
For the aluminum target, data were taken from Refs.~\cite{
Michel1997_ResidualNuclides,
Sisterson1997_LunarRocks,
Dittrich1990_Be10Al26_2,
Steyn1990_Na22Al,
Titarenko2011_AlMonitor,
Leya1998_NobleGas,
Khandaker2011_AlNa,
kuznetsova1962}.
For the copper target, data were taken from Refs.~\cite{
AlSaleh2006_NatCu,
Aleksandrov1987_NatCuNi,
Mills1992_NatCu,
Shahid2015_NatCu,
Garrido2016_NatTiNiCu,
Graves2016_FeCuAl,
Cervenak2020_NatTiCu,
Fox2021_Arsenic}.

Tables~A2--A5 summarize the reference proton-induced fragment production cross sections employed for comparison with the QMD simulations, where ``--" indicates that no reliable reference value was available at the corresponding energy.
These reference values were extracted at selected incident proton energies from fitted experimental energy dependences constructed from the compiled datasets. The corresponding experimental datasets and fitted interpolation curves are shown in Figs.~S1.

\begin{table}[tb]
\centering
\begin{minipage}{0.95\linewidth}
\captionsetup{justification=raggedright,singlelinecheck=false}
\caption{
Experimental fragmentation datasets employed for parameter optimization and additional consistency checks. PCS-LA denotes fragment production cross sections measured within a finite acceptance angle, whereas DCS and DDCS denote differential and double-differential cross sections, respectively.
}
\label{tab:expdata}
\begin{tabular}{lllll}
\hline
Projectile & Energy [MeV/u] & Target(s) & Observable & Ref. \\
\hline
C  & 50  & H, C, O, Al, Ti & DCS, DDCS  & \cite{Divay2017} \\
C  & 95  & H, C, O, Al, Ti & DCS, DDCS  & \cite{Dudouet2013} \\
C  & 290, 400 & H, C, Al, Cu & PCS-LA & \cite{Zeitlin2007} \\
N  & 290, 400 & H, C, Al, Cu & PCS-LA & \cite{Zeitlin2011} \\
O  & 290, 400 & H, C, Al, Cu & PCS-LA & \cite{Zeitlin2011} \\
O  & 400 & C & DCS & \cite{Ridolfi2025} \\
Ne & 290, 400 & H, C, Al, Cu & PCS-LA & \cite{Zeitlin2011} \\
Mg & 400 & H, C, Al, Cu & PCS-LA & \cite{Zeitlin2011} \\
\hline
\end{tabular}
\vspace{1mm}\\
\footnotesize
For C-ion beam datasets at 290 and 400 MeV/u, fragmentation yields were compared for three acceptance-angle conditions and for effective fragment charges of $Z_{\mathrm{eff}} = 5, 4.4, 4, 3.5, 3, 2,$ and $1$, together with charge-changing (CC) cross sections. 
For N-, O-, Ne-, and Mg-ion beam datasets, only large-acceptance measurements were employed; therefore, fragments with atomic numbers $\ge Z_p/2$ were included in the comparison. CC cross sections were also included for these datasets.
H denotes hydrogen-target measurements reported in the original experimental datasets.
\end{minipage}
\end{table}
\begin{table}[htbp]
\centering
\begin{minipage}{0.95\linewidth}
\captionsetup{justification=raggedright,singlelinecheck=false}
\caption{
Reference proton-induced fragment production cross sections for the carbon target extracted from fitted experimental energy dependences. 
Production cross sections are given in barn.
}
\label{tab:protonfitC}
\begin{tabular}{llllllll}
\hline
$E$ [MeV]
& $^{10}$B
& $^{11}$B
& $^{10}$Be
& $^{7}$Be
& $^{9}$Be
& $^{10}$C
& $^{11}$C \\
\hline

30
& --
& --
& --
& 0.0028
& 0.00034
& 0.0011
& 0.0716 \\

50
& 0.0199
& 0.0470
& --
& 0.0198
& 0.0029
& 0.0027
& 0.0829 \\

100
& 0.0182
& 0.0415
& --
& 0.0140
& 0.0028
& 0.0027
& 0.0634 \\

150
& 0.0166
& 0.0365
& 0.0011
& 0.0108
& 0.0032
& 0.0025
& 0.0516 \\

250
& 0.0137
& 0.0282
& 0.0021
& 0.0064
& 0.0040
& 0.0020
& 0.0344 \\
\hline
\end{tabular}
\vspace{1mm}

\footnotesize
Original data are from Refs.~\cite{
Dickson1951_Be7_C,
Brun1962_ProtonBeamIntensity,
Gauvin1962_AlphaSpallation,
Honda1964_Spallation,
Measday1966_C11,
Davids1970_LiBeB_C12,
Fontes1971_LiBe_C12,
Fontes1977_B10B11_C12,
Vdovin1979_50MeVProtonsCNO,
Aleksandrov1990_Be7LightNuclei,
Akagi2013_PositronEmitters,
Matsushita2016_TargetFragmentationCarbon,
Horst2019_PETCrossSections}.
For $^{10}$B and $^{11}$B, cumulative production cross sections were reported in the experimental datasets. 
Therefore, estimated contributions from $^{10}$C and $^{11}$C were subtracted from the cumulative yields prior to comparison with the simulated fragment production cross sections.
\end{minipage}
\end{table}
\begin{table}[htbp]
\centering
\begin{minipage}{0.95\linewidth}
\captionsetup{justification=raggedright,singlelinecheck=false}
\caption{
Reference proton-induced fragment production cross sections for the oxygen target extracted from fitted experimental energy dependences. 
Production cross sections are given in barn.
}
\label{tab:protonfitO}

\begin{tabular}{llllll}
\hline
$E$ [MeV]
& $^{10}$Be
& $^{7}$Be
& $^{11}$C
& $^{13}$N
& $^{15}$O \\
\hline

30
& --
& 0.000429
& 0.00346
& 0.00254
& 0.0642 \\

50
& 0.000072
& 0.00735
& 0.0221
& 0.00596
& 0.0666 \\

100
& 0.000416
& 0.00729
& 0.0100
& 0.00586
& 0.0530 \\

150
& 0.000547
& 0.00743
& 0.00998
& 0.00586
& 0.0420 \\

250
& 0.000950
& 0.00774
& 0.00998
& 0.00586
& 0.0260 \\

\hline
\end{tabular}

\vspace{1mm}
\footnotesize
Original data are from Refs.~\cite{
Dittrich1990_Be10Al26,
Sisterson1997_LunarRocks,
Aleksandrov1990_Be7LightNuclei,
Michel1997_ResidualNuclides,
Bimbot1971_SpallationLightNuclei,
Valentin1963_LightNuclei155MeV,
Foley1962_GammaRadiationO16F19,
Valentin1965_MediumEnergyLightNuclei,
Akagi2013_PositronEmitters,
Masuda2018_CherenkovCrossSections,
Horst2019_PETCrossSections}.
\end{minipage}
\end{table}
\begin{table}[h]
\centering
\begin{minipage}{0.95\linewidth}
\captionsetup{justification=raggedright,singlelinecheck=false}
\caption{
Reference proton-induced fragment production cross sections for the aluminum target extracted from fitted experimental energy dependences. 
Production cross sections are given in barn.
}
\label{tab:protonfitAl}

\begin{tabular}{llllll}
\hline
$E$ [MeV]
& $^{26}$Al
& $^{27}$Mg
& $^{22}$Na
& $^{24}$Na
& $^{22}$Ne \\
\hline

30
& 0.151
& --
& 0.00458
& 0.000210
& -- \\

50
& 0.0758
& --
& 0.0407
& 0.00580
& 0.00963 \\

100
& 0.0496
& 0.000071
& 0.0194
& 0.0117
& 0.0116 \\

150
& 0.0353
& 0.000083
& 0.0186
& 0.0117
& 0.0118 \\

250
& 0.0239
& 0.000114
& 0.0172
& 0.0117
& 0.0121 \\

\hline
\end{tabular}

\vspace{1mm}
\footnotesize
Original data are from Refs.~\cite{
Michel1997_ResidualNuclides,
Sisterson1997_LunarRocks,
Dittrich1990_Be10Al26_2,
Steyn1990_Na22Al,
Titarenko2011_AlMonitor,
Leya1998_NobleGas,
Khandaker2011_AlNa,
kuznetsova1962}.
\end{minipage}
\end{table}
\begin{table}[h]
\centering
\begin{minipage}{0.95\linewidth}
\captionsetup{justification=raggedright,singlelinecheck=false}
\caption{
Reference proton-induced fragment production cross sections for the copper target extracted from fitted experimental energy dependences. 
Production cross sections are given in barn.
}
\label{tab:protonfitCu}

\begin{tabular}{lllll}
\hline
$E$ [MeV]
& $^{60}$Co
& $^{61}$Cu
& $^{57}$Ni
& $^{63}$Zn \\
\hline

30
& 0.00188
& 0.106
& --
& 0.0394 \\

50
& 0.0100
& 0.0828
& 0.00147
& 0.0233 \\

100
& 0.0117
& 0.0431
& --
& 0.00669 \\

150
& 0.0124
& 0.0258
& --
& 0.00490 \\

250
& 0.0140
& 0.00925
& --
& 0.00262 \\

\hline
\end{tabular}

\vspace{1mm}

\footnotesize
Original data are from Refs.~\cite{
AlSaleh2006_NatCu,
Aleksandrov1987_NatCuNi,
Mills1992_NatCu,
Shahid2015_NatCu,
Garrido2016_NatTiNiCu,
Graves2016_FeCuAl,
Cervenak2020_NatTiCu,
Fox2021_Arsenic}.
\end{minipage}
\end{table}

\clearpage

\bibliographystyle{pmbharvard}
\bibliography{references}


\end{document}